\documentstyle[epsf,prl,aps,epsf]{revtex}
\begin{document}
\newcommand{\gsim}{\gtrsim}
\newcommand{\lsim}{\lesssim}
\newcommand{\psim}{\mbox{\raisebox{-1.0ex}{$~\stackrel{\textstyle \propto}
{\textstyle \sim}~$ }}}
\newcommand{\vect}[1]{\mbox{\boldmath${#1}$}}
\newcommand{\lmk}{\left(}
\newcommand{\rmk}{\right)}
\newcommand{\lnk}{\left\{ }
\newcommand{\nn}{\nonumber}
\newcommand{\rnk}{\right\} }
\newcommand{\lkk}{\left[}
\newcommand{\rkk}{\right]}
\newcommand{\lla}{\left\langle}
\newcommand{\p}{\partial}
\newcommand{\rra}{\right\rangle}
\newcommand{\beq}{\begin{equation}}
\newcommand{\eeq}{\end{equation}}
\newcommand{\beqa}{\begin{eqnarray}}
\newcommand{\eeqa}{\end{eqnarray}}
\newcommand{\lab}{\label}
\newcommand{\sol}{M_\odot}

\draft
\title{Possibility of Direct Measurement of the  Acceleration of the
Universe using 0.1Hz Band  Laser Interferometer Gravitational Wave Antenna in Space}

\author{
Naoki Seto$^{1}$,
Seiji Kawamura$^{2}$
and
Takashi Nakamura$^{3}$
}
\address{$^{1}$Department of Earth and Space Science, Osaka
University, Toyonaka 560-0043, Japan}
\address{
$^{2}$National Astronomical Observatory, Mitaka 181-8588, Japan
}
\address{
$^{3}$
Yukawa Institute for Theoretical Physics, Kyoto University,
Kyoto 606-8502, Japan
}

\maketitle
\begin{abstract}

 It may be possible to construct a laser interferometer 
gravitational wave antenna in space with  $h_{rms}\sim 10^{-27}$ at $ f\sim 0.1{\rm Hz}$ 
in this century. We show  possible specification of this
antenna which we call  DECIGO. Using this antenna we show that 1)
 typically  $ 10^5$ ($10^4\sim 10^6$) chirp signals of 
coalescing binary neutron stars per year may be detected with S/N
$\sim 10^4$. 2) We can  directly  measure  
the acceleration of the universe by  ten years observation of binary
 neutron stars. 3) 
The stochastic gravitational waves of $\Omega_{GW}\gsim 10^{-20}$
 predicted by the inflation may be detected by 
 correlation analysis for which 
 effects of the recent cosmic acceleration would
 become 
highly  important. Our formula for phase shift due to accelerating
 motion might be also applied for binary sources of LISA.

\end{abstract}

\section{Introduction}
There are at least four methods to detect  gravitational waves.
They are ;1) Resonant type antenna covering $\sim $kHz band; 
2-a) Laser interferometers on the ground covering 10Hz$ \sim$ kHz band;
2-b) Laser interferometers in space like LISA \cite{lisa} covering
$10^{-4} \sim 
10^{-2}$Hz band;.
3) Residuals of pulsar timing covering $\sim 10^{-8}$Hz band; 
4) Doppler tracking of the spacecraft covering $10^{-4} \sim
10^{-2}$Hz band.
It is quite interesting to note that little has been argued on
possible detectors in  $10^{-2} \sim 10$Hz band.
In this Letter we argue in \S 2  possible specification of such a
detector which we call  DECIGO(DECi hertz Interferometer Gravitational
wave Observatory). In \S 3 we argue that the direct measurement of 
the acceleration of the universe is possible using  DECIGO.
\S 4 will be devoted to discussions.

\section{Specification of  DECIGO}

The sensitivity of a space antenna with an arm length of 1/10 of
LISA and yet the same assumption of the technology level, such as
a laser power of 1 W, the optics of 30 cm, etc. will be $4\times10^{-21}
{\rm Hz}^{-1/2}$ around 0.1 Hz in terms of strain, a factor of 10 better
than the planned LISA sensitivity around 0.1 Hz (see also
http://www.physics.montana.edu/maggie for a project named MAGGIE around
this band). The sensitivity
could be improved by a factor of 1000 for the next generation of
a space antenna with more sophisticated technologies such as
implementation of higher-power lasers and larger optics in order
to increase the effective laser power available on the detectors,
and thus to reduce the shot noise. The ultimate sensitivity of a
space antenna  in the far future could be, however,
$3\times10^{-27}$ 
 around 0.1 Hz in terms of strain, assuming the quantum
limit sensitivity for a 100 kg mass and an arm length of 1/10 of
LISA.
We name this detector DECIGO.
 This requires an enormous amount of effective laser power,
and also requires that the other noise sources, such as gravity
gradient noise, thermal noise, practical noise, etc. should be
all suppressed below the quantum noise.
Here we assume that such an antenna may be available by the
end of this century,  although we note that within the next five
years or so NASA will begin serious discussions of a follow-on
to the planned NASA/ESA LISA mission, so DECIGO technology may
be achieved sooner.
 Note here that
when the pioneering efforts to detect the gravitational
waves started in  the last century using resonant type detectors as
well as laser interferometers, 
few people  expected the present 
achievement in resonant type detectors such as IGEC(bar) \cite{igec}
and in laser inteferometers such as TAMA300\cite{ando},
LIGO, GEO600, and VIRGO (for these detectors see \cite{detec}). 
Therefore all the experimentalists and the theorists on gravitational
waves should not be restricted to the present levels of the
detectors. Our point of view in this Letter is 
believing  the  proverb `` Necessity is the mother of the invention''
 so that we argue why a detector like 
DECIGO is necessary to measure some important
parameters in cosmology. 

The sensitivity of DECIGO, which is optimized at 0.1 Hz, is
assumed to be limited only by radiation pressure noise below 0.1
Hz and shot noise above 0.1 Hz. The contributions of the two
noise sources are equal to each other at 0.1Hz, giving the
quantum limit sensitivity at this frequency. The radiation
pressure noise has a frequency dependence of $\propto f^{-2}$ (in units
of ${\rm Hz}^{-1/2}$) because of
the 
inertia of the mass, while the shot noise has a dependence of
approximately $\propto f^1$ (in units
of ${\rm Hz}^{-1/2}$) because of the signal canceling effect due to
the long arm length.
In figure 1 we show sensitivity of various detectors and
characteristic amplitude $h_c$ for a chirping NS-NS binary at $z=1$.
\section{Direct Measurement of the  Acceleration of the Universe} 

Recent distance measurements for high-redshift supernovae suggest that  
the  expansion of our universe is  accelerating \cite{Riess:1998cb}
which means that the equation of the state of
the universe is dominated by ``dark energy'' with $\rho+3p<0$. 
{\it SuperNova / Acceleration Probe} (SNAP, http://lbl.gov) project will
observe $\sim 2000$ Type Ia supernovae per year up to the redshift
$z\sim 1.7$   so that we may get the accurate luminosity
distance $d_L(z)$ in near future.
Gravitational wave would be also a powerful tool to determine 
$d_L(z)$ \cite{Schutz:1989yw}.

{}From  accurate $d_L(z)$ one may think that it is   possible to determine   the
energy density $\rho(z)$ and the pressure $p(z)$ as   functions of the
redshift.
  However as shown by  Weinberg \cite{Weinberg} and Nakamura \& Chiba \cite{Nakamura:1999mt},
 $\rho(z)$ and $p(z)$ 
can not be determined uniquely from $d_L(z)$ but they 
depend on one free parameter $\Omega_{k0}$ (the spatial curvature). 


Recent measurement of the first peak of the anisotropy of CMB is consistent with a
 flat universe ($\Omega_{k0}=0$) for primordially scale-invariant spectrum predicted by
slow-roll inflation \cite{deBernardis:2000gy} under the
assumption of   $\Lambda$ cosmology.
However  it is  important to determine the curvature of the universe
irrespective of the theoretical assumption on the equation of the
state and the primordial spectra also. In other words an independent
determination of $\Omega_{k0}$ is indispensable since $\Omega_{k0}$ is 
by far the important parameter.
  As discussed in \cite{Nakamura:1999mt}, the direct measurement
 of the cosmic acceleration \cite{Loeb:1998} can be used for this purpose.  Here we point
out that the gravitational waves from the coalescing binary neutron stars at $z\sim 1$ observed by
DECIGO may be used to determine $\Omega_{k0}$.
 Even in the worst case the redundancy
 is important to confirm such an important finding as the dark energy.

\subsection{Cosmic Acceleration}
We consider the propagation of gravitational wave in our isotropic
and homogeneous universe. The   metric is given by
$ ds^2=- dt^2+a(t)^2(dx^2+r(x)^2(d\theta^2+\sin^2\theta d\phi^2))$, 
where $a(t)$ is the scale factor and
 $a(t)r(x)$ represents the angular distance.
The relation between the observed time of the gravitational waves $t_o$ 
at $x=0$ and the emitted time $t_e$  at  the fixed comoving coordinate $x$ is given by
$
\int^{ t_o}_{t_e}\frac{dt}{a(t)}=x=const.
$
Then we have ${dt_o}/{dt_e}={a_o}/{a_e}=(1+z)$ and
\beq
\frac{d^2t_o}{dt_e^2}=(1+z)a_e^{-1}(\p_t a(t_o)-\p_t a(t_e))\equiv
g_{cos}(z)=(1+z)((1+z)H_0-H(z)), \lab{2nd}
\eeq
where $H(z)$ is the Hubble parameter at the redshift $z$ and $H_0$ is
the present Hubble parameter.
For an emitter at the cosmological distance $z\gsim 1$ we have
$g_{cos}(z)\sim O(t_0^{-1})$ where $t_0$ is the age of universe $t_0\sim
3\times 10^{17}$ sec. 
{}From above equations we have 
$
\Delta t_o=\Delta t_e (1+z)+\frac{g_{cos}(z)}2  \Delta t_e^2+\cdots,
$
where   $\Delta t_o$  and $\Delta t_e$ are the arrival time at the
observer and the time at the emitter, respectively.
When we observe the gravitational waves  from the cosmological
distance,  we have
$
\Delta t_o =\Delta T+ X(z) \Delta T^2+\cdots , 
$ with
$
X(z)\equiv g_{cos}(z)/2(1+z)^2,
$
where  $\Delta T=(1+z)\Delta t_e$  is the arrival time  neglecting the
cosmic acceleration/deceleration (the second term). 
Now for $\Delta T\sim 10^9$ sec,  the time lag of the arrival time  due to the cosmic
acceleration/deceleration  amounts to the order of second $\sim 10^{18}/(3\times
10^{17})\sim$ 1 [sec]. From Eq. (\ref{2nd}), if  $X(z)$
is positive,  then $\p_t a(t_o) > \p_t a(t_e)$. This clearly  means that 
our universe is accelerating.  Therefore the value of this time lag  is the direct evidence for the
acceleration/deceleration of the universe.

As shown in \cite{Nakamura:1999mt}, if the accurate value
of $X(z)$ at a single point $z_s$ is available it is  possible  to
determine $\Omega_{k0}$ as
$
\Omega_{k0}=\{1-(dr(z_s)/dz)^2(1+z_s)^2(H_0-2X(z_s))^2\}\{r(z_s)^2H_0^2\}^{-1}, 
$
where we have assumed that the quantity $r(z)\equiv d_L(z)/(1+z)$ is
obtained  accurately, {\it e.g.}, by SNAP. 
Even if the accurate values of $X(z)$ are not available for any points,
we may apply the maximal likelihood method to determine $\Omega_{k0}$.
Using the value of $\Omega_{k0}$ thus determined, we can obtain the equation of state of our universe 
 without any theoretical assumption on its matter content  \cite{Nakamura:1999mt}.

\subsection{Evolution of Phase of Gravitational
Waves from    Coalescing Binary at Cosmological Distance} 
Let us study an inspiraling compact binary system that evolves
secularly by radiating gravitational wave \cite{Cutler:1994ys}. For
simplicity we study 
a circular orbit  and evaluate the gravitational wave  amplitude  and
the energy loss rate by  Newtonian quadrupole formula. 
We basically follow analysis of Cutler \& Flanagan \cite{Cutler:1994ys}
but properly take into account of effects of accelerating motion.
The Fourier transform $\tilde{h} (f)=\int^{\infty}_{-\infty}e^{2\pi i f
t}h(t)dt$  
 for the wave $h(t)$ is evaluated using  the stationary phase
approximation  as
$
\tilde{h}(f)=K d_L(z)^{-1} M_c^{5/6} f^{-7/6} \exp[i\Phi(f)], 
$
where  $K$ is determined by the angular position and the orientation of
the binary relative to the detector, and $M_c$ is the chirp mass of the
system.
Keeping the first order term of the coefficient $X(z)$,  the phase
$\Phi(f)$ of the 
gravitational wave becomes 
\beq
\Phi(f)=2\pi f t_c-\phi_c-\frac{\pi}4+\frac34 (8\pi M_{cz}
f)^{-5/3}-\frac{25}{32768} X(z) f^{-13/3} M_{cz}^{-10/3}
\pi^{-13/3}, \lab{phase2} 
\eeq
where $t_c$ and $\phi_c$ are integral constants and $M_{cz}=M_c(1+z)$ is
the redshifted chirp mass.

If we include the  post-Newtonian (PN) effects  up to
P${}^{1.5}$N-order, the term  $ 3/4 (8\pi M_{cz} f)^{-5/3}$ in
eq.(\ref{phase2}) 
should be  modified    as 
$
 \frac34 (8\pi M_{cz} f)^{-5/3} \lkk 1+\frac{20}9
\lmk\frac{743}{336}+\frac{11\mu}{4 M_c} \rmk x+(4\beta-16\pi) x^{3/2}+\cdots  \rkk, 
$
where $x\equiv (\pi M_t f(1+z))^{2/3}=O(v^2/c^2)$ is  the PN
expansion parameter with $M_t$ being the  total mass of the binaries.
 The term proportional to $\beta$ in P${}^{1.5}$N order ($\propto
x^{3/2}$) is caused by 
the spin effect  \cite{Cutler:1994ys,Cutler:1993tc}. In general
P${}^{N}$N contribution depends on the 
frequency $f$ as $O(f^{(-5+2N)/3})$ and is largely different from the
dependence $f^{-13/3}$ caused by the   cosmic acceleration. 
This difference is very preferable
for the actual signal analysis.

\subsection{the estimation error}
For the circular orbit of the binary neutron stars (NSs) of mass $M_1$ and
$M_2$  
with the separation $a$ at the redshift $z$, the frequency of the gravitational waves $f$
is given by $f=0.1{\rm Hz}(1+z)^{-1}(M_t/2.8\sol)^{1/2}(a/15500{\rm
km})^{-3/2}$. The coalescing time $t_c$, the number of cycles  $N_{cycle}$ and 
the characteristic amplitude of the  waves $h_c$ are given by
\begin{eqnarray}
t_c&=&7(1+z)(M_1/1.4\sol)^{-1}(M_2/1.4\sol)^{-1}(M_t/2.8\sol)^{-1}(a/15500{\rm km })^{4}{\rm yr}\\
N_{cycle}&=&1.66\times 10^7 (M_1/1.4\sol)^{-1}(M_2/1.4\sol)^{-1}(M_t/2.8\sol)^{-1/2}(a/15500{\rm  km})^{5/2}\\
h_c&=&1.45\times 10^{-23}(1+z)^{5/6}(M_c/1.2\sol)^{5/6}(f/0.1{\rm  Hz})^{-1/6}(d_L/10{\rm Gpc  })^{-1}
\end{eqnarray}

Let us evaluate how accurately we can fit the parameter $X(z)$. We take
six parameters $\lambda_i=\{A,M_{cz}, \mu_z,t_c,\phi_c,
M_{cz}^{-10/3}X(z)\}$ 
in the matched 
filtering analysis up to 1PN-order for the phase $\Phi(f)$ and Newtonian
order for the amplitude \cite{Cutler:1994ys}. Here $A$ is the amplitude of
signal $K d_L(z)^{-1} 
M_c^{5/6}$ in the previous subsection and $\mu_z$ is the redshifted reduced mass $\mu_z=(1+z)M_1M_2/M_t$. As the chirp mass $M_{cz}$ can be determined quite
accurately, we simply put
$\Delta
X(z)=\Delta\lnk 
M_{cz}^{-10/3}X(z)\rnk/M_{cz}^{-10/3}$.  
For simplicity we fix the redshift of sources at $z=1$ and calculate S/N and the
 error $\Delta X$ for equal mass binaries  with
various integration time $\Delta t$ before 
 coalescence. We use the effective factor $1/\sqrt5$ for reduction of
antenna sensitivity due to its rotation \cite{lisa}.
 For the present analysis we 
neglect  the binary
confusion noise since double White Dwarf binaries do not exist at
frequency  $f\gsim
0.1$Hz 
\cite{hils}.

 We found that 
we can detect NS-NS binaries at $z=1$ with  $S/N\simeq 20000$  and $\Delta
X/t_0^{-1}\simeq 7.0\times 10^{-3}$ for integration time $\Delta t=16$yr
($N_{cycle}\sim10^7$  
orbital 
cycles),  and  $S/N\simeq 10000$ and   $\Delta
X/t_0^{-1}\simeq 1.26$  for $\Delta t=1$yr.  With this  detector 
it would be possible to 
determine $X(z)$   and  obtain
the information of the cosmic acceleration quite accurately. 
With $\Delta T=16$yr we have the estimation error for the redshifted masses as 
$\Delta M_{cz}/M_{cz}=1.5\times 10^{-11}$, $\Delta
\mu_{z}/\mu_{z}=4.2\times 10^{-8}$ and for the wave amplitude $\Delta
A/A\sim(S/N)^{-1}=5\times 10^{-5} $.  Although  the more detailed
study is needed to estimate the  error of  the binary inclination
angle, it is expected that the luminosity distance $d_L$ can be
determined accurately
so that  the redshift $z$ can be determined  using the inverse function 
$z=d_L^{-1}$(distance) of the accurate luminosity distance
 from {\it e.g.} SNAP. 
As a result we can know two (not redshifted) masses $M_1$ and  $M_2$
for $\sim 10^5$ binaries per year up to $z=1$ \cite{Kalogera:2000dz}.
This  will be large enough to establish the  mass
function of NS which would bring us important implications
for the equation of the state  of the high density matter and the
explosion 
mechanisms of TypeII supernovae.

As the S/N and the estimation error scale as
$S/N\propto h_{rms}^{-1}$ and $ \Delta X\propto h_{rms}$,
we can 
attain $\Delta
X/t_0^{-1}\simeq 7.0$ for the integration time $T=16$yr using a less sensitive
detector with $h_{rms}\sim
10^{-24}$ (1000 times worse).  Even though the error bar $\Delta X$ is
fairly large 
for this detector,
the likelihood analysis 
would be an  efficient  approach to study the cosmic acceleration.
Considering the estimated  
cosmological coalescence  rate 
of NS-NS binaries ($\gsim 2\times(10{\rm Gpc}/350{\rm Mpc})^3\sim 10^5{\rm
yr}^{-1}$) 
\cite{Kalogera:2000dz}, we may expect the decrease of the estimation
error $\Delta X$ roughly  by a
factor of  $\sim 1/300=1/\sqrt{10^5}$.


\subsection{Acceleration in the Very Early Universe}

In the inflationary phase there was an extremely  rapid  acceleration of the 
universe. In this phase the gravitational waves were generated by 
quantum  fluctuation  \cite{Maggiore:2000vm}. 
With CMB quadrupole anisotropies measured by COBE, the
slow-roll inflation
model predicts a  constraint on the stochastic background
$\Omega_{GW}\lsim 10^{-15}- 
10^{-16}$ at $f\sim 0.1$Hz  \cite{Turner:1997ck}.  
Ungarelli and Vecchio \cite{Ungarelli:2001jp} discussed that 
the strain  sensitivity $h_{rms}\sim  10^{-24}$   is the required level at  $f\sim 0.1$Hz for
detecting $\Omega_{GW}\sim 10^{-16}$ by correlating
two detectors  for  decades (see also Ref.\cite{cornish}).   It is
important to note that  the band  $f> 0.1$ Hz is free from stochastic
backgrounds generated by White Dwarf binaries.  The radiation from
neutron stars binaries is present in this band and it is indispensable
 to remove 
their contributions accurately from data stream, where effects of the
cosmic acceleration would be highly important.  Thus measurement of the
present-day cosmic acceleration is closely related to detection of the
primordial gravitational wave background that is one of the most
interesting targets in cosmology.  
If DECIGO with $h_{rms}\sim 2\times 10^{-27}$ at $f\sim 0.1$Hz
is available  we can detect  the   primordial gravitational waves
background even if the energy density is extremely low  $\Omega_{GW}\sim 10^{-20}$  by correlating two
detectors for a decade.  

Confusion noise due to NS-NS (or NS-BH, BH-BH) binaries
 might be important in the band $f\sim 0.1$Hz.
 Ungarelli and Vecchio \cite{Ungarelli:2001jp} 
investigated  the critical frequency $f_g$ where we
 can, in principle, remove signal from individual NS-NS 
binaries by matched filtering analysis and the observed
 window becomes transparent to the primordial stochastic
 background.  They roughly estimated
$f_g\sim 0.1$Hz where the number of binaries per
 frequency bin ($\sim 10^{-8}$Hz) is less than one. 
 But binaries around $f\sim f_g\sim 0.1$ 
chirp significantly  within observing time scale  
and  the situation would be  more complicated   than  
  monochromatic sources \cite{hils}.
Although more detailed analysis is needed, 
 a much
 smaller NS-NS coalescence rate than
$\sim 10^5{\rm yr^{-1}}$ might be
required for our analysis to be valid.

\section{Discussions}
The determination of the   angular position of the source is crucial for matching the
 phase \cite{lisa}.   The phase 
modulation at the orbital radius 1AU corresponds to 2AU/$c\sim 1000$[sec].
Thus,  in order to match the phase within the accuracy of $0.1$[sec] we
need to determine the  angular position  with precision $\sim
0.1/1000 ~[\rm  rad]\sim 20''$. 
In the matched filtering analysis we can simultaneously
 fit parameters of  the
angular position  as well as the relative acceleration between the source and the
barycenter of the solar system.   Due to their
correlation in the Fisher matrix,  the measured acceleration would be
somewhat degraded if we cannot determine  the angular position  by other
observational methods. Using the gravitational wave alone, we can, in advance, specify the
coalescence time  and the angular position of the source within some error box. 
If coalescence of NS-NS binaries would release the optical signal ({\it
e.g.} Gamma Ray Bursts as proposed by 
\cite{Paczynski:1986}) we may measure  the angular position accurately  by
pointing  telescopes toward   the error box at  the expected
coalescence time from the chirp signal.
Therefore  we have not tried  to fit  the angular position of the source in the matched filtering method
\cite{lisa}. 
 We might also
determine the redshift of the source by using optical information of host
galaxies.

Let us discuss the effects of the local motion  $g_{local}$ of the emitteron the
second derivative 
$d^2t_o/dt_e^2$ (see {\it e.g.} Ref.\cite{Damour:1991}). 
As the effect of bulk motion of galaxy is much smaller than cosmological
effect,  we estimate the internal acceleration within the  galaxy 
based on the observational result of NS-NS binary PSR 1913+16.  As shown in Table
1 of \cite{Damour:1991},  the dominant contribution of its acceleration 
$\ddot{x}$ comes from the  global Galactic potential field and has time
scale $c/\ddot{x}\sim10t_0(R_e/10{\rm kpc})(V_{rot}/200{\rm km s^{-1}})$
that can be comparable to the cosmic signal $g_{cos}$ where $R_e$ 
is the effective radius of the acceleration and $V_{rot}$ is galactic
rotation velocity.
However  the contamination of  local effect $g_{local}$ can be reduced by taking
the statistical average of many  binaries as  $\lla g_{cos}+g_{local}\rra=\lla g_{cos}\rra$. 
 We also note that the
cosmological change in phase of a coalescing binary [given by
the last term in Eq. (3)]  may have other
applications, and may under certain circumstances be observable
by the planned LISA mission.

In conclusion we would like to encourage the further design study
of DECIGO and the theoretical study of the sources of gravitational
 waves 
for DECIGO. Even if we may not see the construction of DECIGO
 in our life since the highly advanced technology is needed,
we are sure that our children or grandchildren will decide and go DECIGO.
 

We would like to thank  anonymous  referees including 
helpful comments  on the construction time of DECIGO. 
N.S. would like to thank H. Tagoshi for useful discussions and
 M. Sasaki for his invaluable advice.
This work was supported in part by
Grant-in-Aid of Monbu-Kagaku-Shyo Nos.11640274, 09NP0801 and 0001416.

\begin{figure}[h]
 \begin{center}
 \epsfxsize=8.cm
 \begin{minipage}{\epsfxsize} \epsffile{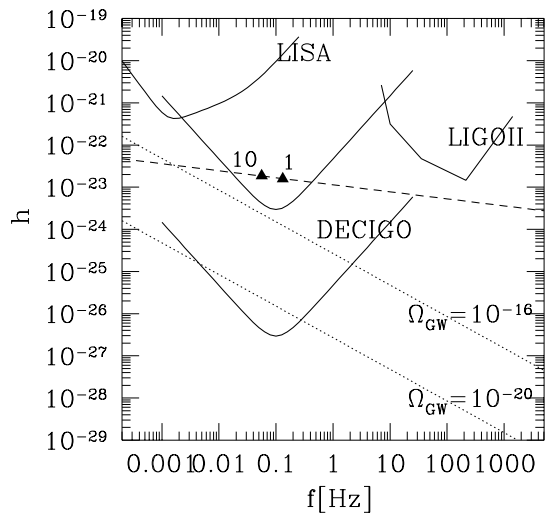} \end{minipage}
 \end{center}
\caption[]{Sensitivity (effectively S/N=1) for various detectors (LISA,
 DECIGO, LIGOII and a detector $10^3$  times less sensitive than DECIGO)
 in 
 the 
 form of $h_{rms}$ (solid lines). 
The dashed line represents  evolution of the characteristic amplitude
 $h_c$ for NS-NS binary at $z=1$ (filled triangles; wave frequencies at
 1yr and 10 yr before coalescence). The dotted lines represent the
 required sensitivity for detecting stochastic background with
 $\Omega_{GW}=10^{-16}$ and $\Omega_{GW}=10^{-20}$ by ten years
 correlation analysis (S/N=1).}
\end{figure}

\end{document}